\documentclass[12pt]{article}
\usepackage{amsmath}
\usepackage{graphicx,psfrag,epsf}
\usepackage{enumerate}
\usepackage{natbib}
\usepackage{url} 

\usepackage{amssymb, amsfonts, amscd, xspace, pifont, amsthm}
\usepackage{mathrsfs}
\usepackage{booktabs}
\usepackage{multirow}
\usepackage{epstopdf}
\usepackage{color}
\usepackage[lined,boxed,commentsnumbered]{algorithm2e}
\usepackage{tikz}
\usetikzlibrary{fit,positioning,arrows,automata}

\newcommand{\blind}{0}

\newtheorem{prop}{Proposition}[section]

\numberwithin{equation}{section}
\def\twoImages#1#2#3#4#5#6 
{
  \centerline{\hfill\makebox[#2]{#3}\hfill\makebox[#5]{#6}\hfill}
  \centerline{\hfill
    \includegraphics[width=#2]{#1}
    \hfill
    \includegraphics[width=#5]{#4}
    \hfill}
}

\newenvironment{remark}[1][Remark]{\begin{trivlist}
\item[\hskip \labelsep {\bfseries #1}]}{\end{trivlist}}

\DeclareMathOperator*{\argmax}{arg\,max}

\begin{document}

\def\spacingset#1{\renewcommand{\baselinestretch}%
{#1}\small\normalsize} \spacingset{1}

\if0\blind
{
  \title{Network Inference from Temporal-Dependent Grouped Observations}
  \author{Yunpeng Zhao\thanks{
    The author gratefully acknowledges NSF DMS 1513004. }\hspace{.2cm}\\
    School of Mathematical and Natural Sciences, Arizona State University\\
    }
  \maketitle
} \fi

\if1\blind
{
  \bigskip
  \bigskip
  \bigskip
  \begin{center}
    {\LARGE\bf Network Inference from Temporal-Dependent Grouped Observations}
\end{center}
  \medskip
} \fi

\date{}
\bigskip
\begin{abstract}
In social network analysis, the observed data is usually some social behavior, such as the formation of groups, rather than an explicit network structure. Zhao and Weko (2017) propose a model-based approach called the hub model to infer implicit networks from grouped observations. The hub model assumes independence between groups, which sometimes is not valid in practice. In this article, we generalize the idea of the hub model into the case of grouped observations with temporal dependence. As in the hub model, we assume that the group at each time point is gathered by one leader. Unlike in the hub model, the group leaders are not sampled independently but follow a Markov chain, and other members in adjacent groups can also be correlated. 

An expectation-maximization (EM) algorithm is developed for this model and a polynomial-time algorithm is proposed for the E-step. The performance of the new model is evaluated under different simulation settings. We apply this model to a data set of the Kibale Chimpanzee Project. 

\end{abstract}

\noindent%
\vfill

\newpage
\spacingset{1.45} 

\section{Introduction}

A network is a data structure consisting of nodes (vertices) connected by links (edges). A network with $n$ nodes can be represented by an $n \times n$ adjacency matrix $A=[A_{ij}]$, where $A_{ij}>0$ if there is an edge between nodes $i$ and $j$ and $A_{ij}=0$ otherwise. A network $A$ can be weighted, where $A_{ij}$ measures the link strength between node $i$ and $j$. Network analysis has drawn increasing attention in a number of fields such as social sciences \citep{Wasserman94,liben2007link}, physics \citep{Barabasi&Albert1999,Newman&Girvan2004}, computer science \citep{Getoor2005}, biology \citep{stark2006biogrid} and statistics \citep{Bickel&Chen2009,hunter2006inference}.

Traditionally, statistical network analysis deals with inferences concerning parameters of an observed network, i.e., an observed adjacency matrix $A$ (see \citet{Newman2010, Goldenberg2010, zhao2017survey} for reviews of models and techniques for analyzing observed networks). 

In this article, we focus on the case that the network is unobserved and to be estimated. What we do observe is a collection of subsets of nodes. Each subset is called a \textit{group} by \citet{Zhao2015} and a data set consisting of such groups is referred to as \textit{grouped data}. We continue to use the term grouped data in this article. 


To better explain the structure of grouped data, we introduce some notations. For a set of $n$ individuals, $V=\{v_1,..., v_n\}$, we observe $T$ subsets $V^{1},...,V^{T}$ at times $1,...,T$, called groups. 
 Each observed subset $V^{t}$ can be represented as an $n$ length row vector $G^{t}$, where
\[ G_i^{t} = \left\{ 
   \begin{array}{l l}
     1 & \quad \textnormal{if $v_i\in V^{t}$,}\\
     0 & \quad \textnormal{otherwise. }
   \end{array} \right.\]
Let $G$ be a $T \times n$ matrix with $G^{t}$ being its rows. For simplicity, we will slightly abuse the notation: we will also refer to the indicator vector $G^{t}$ as a group from now on. 

In this article, we analyze the grouped data from the so-called \textit{social network perspective} \citep{Moreno34}. The grouping behavior of the individuals is presumed to be governed by a latent social network $A$. The objective of this article is to infer the latent network from the groups being observed. In other words, we aim to estimate the link strength between individuals using the information about their presence in the groups. 
\begin{remark}
\end{remark}
\begin{enumerate}
	\item Throughout this paper, we consider only the case that there exists one and only one group at a time $t$. 
	\item The groups at different times can overlap. In fact, it is only plausible to make meaningful inferences of $A$ from $G$ if groups overlap. If all groups are disjoint, the best inference is to use a clique, i.e., a fully connected subgraph to estimate the relationships within each group.
\end{enumerate}

\citet{Wasserman94} discuss grouped data (which they called \textit{affiliation networks}) as well as some empirical methods and graphical representations for this type of data in Chapter 8 of \textit{Social Network Analysis: Methods and Applications}. The authors give an illustrative example of six children and three birthday parties (page 299), which is shown in Table \ref{T:Birthday}. By the use of the notation above, $G^2_2=1$ since Drew attended Party 2, but $G^3_2=0$ since Drew did not attend Party 3. 

\begin{table}[h!]
\caption{\small Dataset for six children and three birthday parties. Adapted from Fig. 8.1, page 299 of \citet{Wasserman94}.}
\label{T:Birthday}
\begin{center}
\begin{tabular}{|c  | c c c c c c |}
\hline
Party & Allison & Drew & Eliot & Keith & Ross & Sarah \\
\hline
1 & 1 & 0 & 0 & 0 & 1 & 1\\
2 & 0 & 1 & 1 & 0 & 1 & 1\\
3 & 1 & 0 & 1 & 1 & 1 & 0\\
\hline
\end{tabular}
\end{center}
\end{table} 

Numerous researchers in social sciences have been interested in grouped data, in particular how to infer social structures from such data. \citet{Wasserman94} provide a list of such data sets (pages 295-296). For example, \citet{Galaskiewicz1985} collected CEO membership data that consisted of their participation in clubs, cultural boards and corporate boards of directors (see \citet{Wasserman94} page 755 for the data). As another example (not in \citet{Wasserman94}), \citet{Noordin} collected a data set of 79 terrorists' presence in meetings, trainings and other events. 

Grouped data are also popular in the study of social behaviors of animals. Social network analysis (SNA) has become an important tool in this area \citep{Whitehead2008,croft2011hypothesis}, but direct linkages between animals are often difficult or expensive to record \citep{croft2011hypothesis} and certain methods used for collecting human network data, such as surveys, are clearly impossible. On the contrary, grouped data such as groups of dolphins \citep{Bejder98} or flocks of birds \citep{farine2012} are relatively easier to identify and record.

\citet{Zhao2015} use group data to study novels by treating the characters of a novel appearing in the same paragraph as a group and using the inferred network structure to interpret the relationships between characters. 

Despite the popularity of group data, existing methods for network inference from grouped data are mainly ad-hoc approaches from the social sciences literature. A simple technique is to count the number of times that a pair of nodes appears in the same group. This measure has been called different names by different authors, e.g., the \textit{co-citation} matrix in Section 6.4 of \citet{Newman2010} or the \textit{sociomatrix} in Section 8.4 of \citet{Wasserman94}. \citet{Zhao2015} refer to this measure as the \textit{co-occurrence matrix}. The \textit{half weight index} \citep{Cairns87} is an alternative approach that uses the conditional frequencies of co-occurrences as estimates. A common difficulty of such methods is that they provide no statistical model to connect these descriptive statistics with the latent network.

\citet{Zhao2015} recently proposed a model-based approach for grouped observations. 
In the so-called \textit{hub model}, $G^t$s are modeled as independently and identically distributed random vectors and  there is a central node called \textit{hub} or group leader in each group, who gathers other members into the group. For example, the hub is the child who hosted the party in Table \ref{T:Birthday}.

A crucial assumption made in \citet{Zhao2015} is that the groups are assumed to be independently generated by the hub model. In some cases, this assumption is reasonable if each group forms spontaneously. The assumption can also be approximately satisfied if researchers collect grouped data with sufficiently long time intervals between observations (see \citet{Bejder98} for discussion). 

The independence assumption however may not be valid in other situations. In most practical situations, the grouped observations are temporal-dependent by default. For example, in a study of animal behavior, researchers may observe the behavior of animals on an hourly or daily basis. In Section \ref{sec:data}, we analyze such a data set consisting of groups of wild chimpanzees studied by the Kibale Chimpanzee Project. 
It is inappropriate to assume that every group is independent from the previous group. A more plausible point of view is that the group at a particular time is a transformation of the previous group. That is, some new members may join the group and some may leave, but the group maintains a certain level of stability. 

We generalize the idea of the hub model in order to accommodate temporal dependence between groups. We call the new model the temporal-dependent hub model, or the temporal-dependent model in short. This new model allows for dependency between group leaders as well as between other group members. We explain both dependency assumptions in the next two paragraphs.  

As in the classical hub model, we assume there is one leader for each group. Leaders however are not sampled independently in the temporal-dependent model, but follow a Markov chain. That is, the probability of a certain node being the current leader depends on the leader in the previous group. 

For other group members, we consider the following two cases to make the model flexible enough. If the current leader is inside the previous group, then we treat this group as a transformation of the previous one. If the new leader is from outside the previous group (e.g., some event occurs and completely breaks the previous group) then we treat this group as the start of a new segment. In this case, the leader will select the group members as in the classical hub model, i.e., independently of whether or not they were members of the previous group. 

As shown in Section \ref{sec:alg}, the temporal-dependent hub model can be viewed as a generalization of the hidden Markov model (HMM) when the group leaders are latent. An efficient algorithm is thus developed for model fitting. Furthermore, the temporal-dependent hub model provides estimates of the elements of the adjacency matrix with lower mean squared errors according to numerical studies in Section \ref{sec:simu}. 

Finally, we discuss some related works. First, the temporal-dependent hub model is fundamentally different from many existing models for dynamic networks, such as the preferential attachment model \citep{Barabasi&Albert1999}, discrete/continuous time Markov models \citep{snijders2001statistical,hanneke2007discrete}, etc. In these works, the observed data are snapshots of the network at different time points. In this article, the unknown parameters are a single latent network and the observations are groups with temporal-dependent structures. 

Second, there are recent studies on estimating latent networks or related latent structures in dynamic settings, but from data structures that are different from groups. \citet{guo2015bayesian} propose a Bayesian model to infer latent relationships between people from a special type of data -- the evolution of people's language over time. 
\citet{Robinson2013} propose a latent process model for dynamic relational network data. Such a data set consists of binary interactions at different times. \citet{blundell2012modelling} propose a nonparametric Bayesian approach for estimating latent communities from a similar data type. The grouped data we consider in this article are more complicated than binary interactions in the sense that, unlike a linked pair, the links within a group consisting of more than two members are unknown. 

Third, there are other interesting works on modeling latent social networks from survey data and such data only provide partial information of a latent network. These survey data also have different structures from grouped data.  \citet{mccormick2015latent} propose a latent surface model for aggregated relational data collected by asking
respondents the number of  connections they have with members of a certain subpopulation. In this work, the network structure for the population is latent.
\citet{admiraal2016modeling} fit exponential-family random graph models (ERGMs) to latent heterosexual partnership networks, with degree distributions and mixing totals being sufficient statistics in the exponential family. Those statistics for the underlying population are inferred from cross-sectional survey data. 
 \cite{krivitsky2017inference} fit ERGMs to egocentrically sampled data, which provide information about respondents
and anonymized information on their network neighbors. 

\section{Model}\label{sec:Model}

\subsection{The classical hub model}

We briefly state the generating mechanism of the classical hub model \citep{Zhao2015}.
The hub model assumes one leader for each group. The leader of $G^{t}$ is denoted by $z^t$. 


Under the hub model, each group $G^{t}$ is independently generated by the following two steps. 

\begin{enumerate}
	\item The group leader is sampled from a multinomial distribution with parameter $\rho=(\rho_1,...,\rho_n)$, i.e., $\mathbb{P}(z^t=i)=\rho_i$, with $\sum_i \rho_i=1$.

	\item The group leader, $v_i$, will choose to include $v_j$ in the group with probability $A_{ij}$, i.e., $	\mathbb{P}(G_j^{t}=1|z^t=i)=A_{ij}$.
\end{enumerate}

\subsection{Generating mechanism of the temporal-dependent hub model}\label{generating}
The hub model assumes that all the groups are generated independently across time. In practice, it is more natural to model the groups as temporal-dependent observations.

We first explain the idea of the generating mechanism of temporal-dependent groups and then give the formal definition. We generalize the idea of the hub model into the temporal-dependent setting. Specifically, we assume there is only one leader $z^t$ at each time who brought the group together, but the group at time $t$ depends on the previous group, which is different from the classical hub model.   

At time $t=1$, the group is generated from the classical hub model. For $t=2,...,T$, the group leader $z^t$ can remain the same as the previous leader or change to a new one. We assume that the leader $z^t$ will remain as $z^{t-1}$ with a higher probability than the probability of changing to any other node.  

If the new leader is outside the previous group, then the current group is considered the start of a new segment and is generated by the classical hub model. It is worth noting that technically, the generation of the new group however still depends on the previous group. This will become clearer after we introduce the likelihood function. For the case that the new leader is inside the previous group -- that is, if the leader remains unchanged, or the leader changes but is still a member of the previous group -- we propose the following \textit{In-and-Out procedure}: for any node $v_j$ being in the previous group, it will remain in $G^{t}$ with a probability higher than $A_{z^t,j}$ -- the probability in the classical hub model. On the contrary, for any node $v_k$ not being in the previous group, it will enter $G^{t}$ with a probability lower than $A_{z^t,k}$. Intuitively, this \textit{In-and-Out procedure} assumes that when a group forms, it will maintain a certain level of stability. 

We now give the formal definition of the generating mechanism as follows:
\begin{itemize}

\item Step 1: (Classical hub model). When $t=1$, $G^{t}$ is generated by the following two substeps. 
\begin{itemize}
 \item [1)] The leader is sampled from a multinomial distribution with parameter $\rho=(\rho_1,...,\rho_n)$, i.e., $$ \mathbb{P}(z^{t}=i)=\rho_i \overset{\Delta}{=} \frac{\exp (u_i)}{\sum_{k=1}^n \exp(u_k)}, $$
 where $u_i \in \mathbb{R}$ for $i=1,...,n$. 
 \item [2)] The leader $v_i$ will choose to include $v_j$ in the group with probability $A_{ij}$, i.e., $	\mathbb{P}(G_j^{t}=1|z^{t}=i)=A_{ij}$, where $A_{ii} \equiv 1$ and 
\begin{align*}
A_{ij}= A_{ji}\overset{\Delta}{=}\frac{\exp(\theta_{ij})}{1+\exp( \theta_{ij})}. 
\end{align*}  
Here, $\theta_{ii}=\infty$ for $i=1,...,n$ and $\theta_{ij} \in \overline{\mathbb{R}}$ for $i \neq j$. We allow some $\theta_{ij}$ to be $\pm \infty$ so that the corresponding $A_{ij}$ can be 1 or 0. 
 \end{itemize}
\item Step 2: (Leader change). For $t=2,...,T$, 
\begin{align*}
\mathbb{P}(z^t=i| z^{t-1}) = \frac{\exp (u_i+\alpha I(z^{t-1}=i))}{\sum_{k=1}^n \exp(u_k+\alpha I(z^{t-1}=k))},
\end{align*}
where $\alpha \in \mathbb{R}$. 
\item Step 3: (In-and-Out procedure). For $t=2,...,T$, given $v_i$ being the leader, $G^{t}$ is generated by the following mechanism: 

If $v_i$ is not within $G^{t-1}$, then it will include each $v_j$ in the group with probability $A_{ij}$; otherwise, see below: 
\begin{itemize}
 \item [1)] If $G^{t-1}_j=1$, $v_i$ will include $v_j$ in the group with probability
\begin{align*}
 B_{ij}=B_{ji}\overset{\Delta}{=}  \frac{\exp( \theta_{ij}+\beta )}{1+\exp( \theta_{ij}+\beta )}, 
\end{align*}
where $\beta \in \mathbb{R}$. 
 \item [2)] If $G^{t-1}_j=0$, $v_i$ will include $v_j$ in the group with probability
\begin{align*}
 C_{ij}=C_{ji}\overset{\Delta}{=}  \frac{\exp( \theta_{ij}+\gamma )}{1+\exp( \theta_{ij}+\gamma )}, 
\end{align*}
where $\gamma \in \mathbb{R}$. 
\end{itemize}

\end{itemize}
For clarity of notation, we now give the vector/matrix form. Define $z=(z^{1},..., z^{T})$, $u=(u_1,...,u_n)$ and $\rho=(\rho_1,...,\rho_n)$. Define $\theta=[\theta_{ij}]_{1\leq i \leq n, 1\leq j \leq n}$, $A=[A_{ij}]_{1\leq i \leq n, 1\leq j \leq n}$, $B=[B_{ij}]_{1\leq i \leq n, 1\leq j \leq n}$ and $C=[C_{ij}]_{1\leq i \leq n, 1\leq j \leq n}$. Furthermore, we assume $\theta$, $A$, $B$ and $C$ to be symmetric in order to avoid any issue of identifiability (see the discussion in \cite{Zhao2015}).
\begin{remark}
\end{remark}
\begin{enumerate}
	 \setcounter{enumi}{2}
\item In the definition above, $u_i$ and $\theta_{ij}$ are simply a reparameterization of $\rho_i$ and $A_{ij}$ in exponential form. This is to make optimization more convenient, since log-likelihood is convex under this parametrization. 

\item The parameters $\alpha$, $\beta$ and $\gamma$ characterize the dependency between the groups. $\alpha$ is the adjustment factor, which controls the probability that a leader in the previous group remains as a leader. $\beta$ is the adjustment factor for nodes being inside the previous group. And $\gamma$ is the adjustment factor for nodes being from outside the previous group. We do not enforce $\alpha>0$, $\beta>0$ and $\gamma<0$ in the model fitting. Instead, we test these assumptions for the data example in Section \ref{sec:data}. 

\item The parameters $A$, $\beta$ and $\gamma$ are identifiable. The key observation is that the identifiability of $A$ can simply be obtained by $G^1$ since the first group only depends on $\rho$ and $A$. This is essentially the identifiability of the classical hub model. The proof is given by Theorem 1 in \cite{Zhao2015}, under the condition of $A$ being symmetrical. With the ``baseline'' $A$ being separately identified, the two adjustment factors $\beta$ and $\gamma$ are accordingly identifiable. 

\item The parameters $(u_1,...,u_n)$ are non-identifiable under this parametrization, since $(u_1+\delta,...,u_n+\delta)$ gives the same likelihood. We will discuss the solution to this problem in Section \ref{sec:alg} after introducing the algorithm.

\end{enumerate}

\subsection{Likelihood}
For notational convenience in the likelihood, we indicate the leader in group $G^t$ by an $n$ length vector, $S^{t}$, where
\[ S_i^{t} = \left\{ 
   \begin{array}{l l}
     1 & \quad \text{if $z^t=i$},\\
     0 & \quad \text{otherwise}.
   \end{array} \right.\]   
Only one element of $S^{t}$ is allowed to be 1. $S^{t}$ is simply another representation of $z^{t}$. Let $S$ be a $T \times n$ matrix, with $S^{t}$ being its rows. 

Clearly, $ \{S^1,...,S^T \} $ is a Markov chain according to the generating mechanism. 
Let $\Phi_{ij}=\mathbb{P}(z^t=i|z^{t-1}=j)$ be the transition probability and $\Phi=[\Phi_{ij}]_{n\times n}$. We summarize all introduced notations in Table \ref{Table:notation}.

\begin{table}
\caption{Summary of Notation}
\label{Table:notation}
\begin{tabular}{|c | c| c | c |}
\hline
& Notations & Remark \\
\hline
Parameter  & $ \rho_i$  & Probability of $v_i$ being the leader of $G^1$ \\
			\hline
			& $u_i$ & $\rho_i=\frac{\exp(u_i)}{\sum_{k=1}^n \exp(u_k)} $ \\
			\hline
      & $ \alpha $ & Adjustment factor for remaining leaders \\
      \hline
		  & $\Phi_{ij}$ & $\Phi_{ij}=\mathbb{P}(z^t=i|z^{t-1}=j) $ \\
			\hline
          & $ A_{ij}$ & Probability of $v_j$ being inside the group \\
					&           & when $v_i$ is the leader in a newly formed group \\
			\hline
			    & $ \theta_{ij}$ & $\theta_{ij}=\log \frac{A_{ij}}{1-A_{ij}} $ \\
			\hline
					& $ \beta$  & Adjustment factor for nodes being inside the previous group \\
			\hline
					& $ B_{ij}$ & Adjusted probability of $v_j$ being inside the group \\
					&           & when inside the previous group \\
			\hline 
		   	& $ \gamma$  & Adjustment factor for nodes being outside the previous group \\
			\hline 
				& $ C_{ij}$ & Adjusted probability of $v_j$ being inside the group \\
					&           & when outside the previous group \\
		  \hline
Data    & $ G^{t}$ & Group at time $t$ \\
			\hline
			  & $ z^t $ & Leader at time $t$ \\
			\hline
			  & $ S^{t}$ & Indicator of the leader at time $t$, with only one element being 1 \\
			\hline
Index   & $n$    & Size of the network \\
       \hline 
			  & $T$    & Number of groups (sample size) \\
				\hline 
\end{tabular}

\end{table}

We now give the joint log-likelihood of $S$ and $G$ for the model defined in the previous subsection:
\begin{align}
  & \log \mathbb{P} (S,G|\alpha, \beta, \gamma, \theta, u) \nonumber   \\
= & \sum_{i=1}^n S_i^{1} \log \rho_i +\sum_{t=2}^T \sum_{i=1}^n \sum_{j=1}^n S_i^{t}S_j^{t-1} \log \Phi_{ij} \nonumber \\
  &+ \sum_{i=1}^n \sum_{j=1}^n \left \{  S_i^{1} G_j^{1} \log A_{ij}+S_i^{1} (1-G_j^{1}) \log (1-A_{ij})   \right \} \nonumber \\
  &+ \sum_{t=2}^T \sum_{i=1}^n \sum_{j=1}^n \left \{  S_i^{t} (1-G_i^{t-1}) G_j^{t} \log A_{ij}+S_i^{t}(1-G_i^{t-1}) (1-G_j^{t}) \log (1-A_{ij})   \right \} \nonumber \\
	& + \sum_{t=2}^T \sum_{i=1}^n \sum_{j=1}^n \left \{ S_i^{t}G_i^{t-1} G_j^{t-1} G_j^{t} \log B_{ij}+ S_i^{t}G_i^{t-1} G_j^{t-1} (1-G_j^{t}) \log (1-B_{ij})   \right \} \nonumber \\
		& + \sum_{t=2}^T \sum_{i=1}^n \sum_{j=1}^n \left \{ S_i^{t}G_i^{t-1} (1-G_j^{t-1}) G_j^{t} \log C_{ij}+ S_i^{t}G_i^{t-1} (1-G_j^{t-1}) (1-G_j^{t}) \log (1-C_{ij})   \right \}. \label{joint_lik} 
\end{align}
Note that $\alpha, \beta, \gamma, \theta$ and $u$ are essentially the parameters of this model and $\rho$, $\Phi$, $A$, $B$ and $C$ are their functions. 
Despite its length, Equation \eqref{joint_lik} has a clear structure. The 1st line gives the log-likelihood of $S$. The 2nd line gives the log-likelihood of $G^1$ given $S^1$. The 3rd line gives the log-likelihood of $G^t$ given that the current leader $z^t$ is outside the previous group $G^{t-1}$. The 4th and 5th lines give the log-likelihood of $G^t$ given that $z^t$ is inside $G^{t-1}$, based on the \textit{In-and-Out procedure}. 

Equivalent to \eqref{joint_lik}, we can write the likelihood as a product of conditional probabilities: 
\begin{align*}
\mathbb{P} (S,G) = \mathbb{P} (S^1)  \mathbb{P}(G^1|S^1) \prod_{t=2}^T \mathbb{P}(S^t|S^{t-1})  \prod_{t=2}^T \mathbb{P}(G^t|S^{t},G^{t-1}).
\end{align*}
This factorization can be represented by a Bayesian network (Figure \ref{F:Bayesian}), where a node represents a variable and a directed arc is drawn from node $X$ to node $Y$ if $Y$ is conditioned on $X$ in the factorization. (Refer to \citet{jordan1999introduction} for a comprehensive introduction to Bayesian networks). This Bayesian network should not be confused with the latent network $A$ -- the former is a representation of the dependency structure between variables while the latter reflects the relationships between the group members. 

Furthermore, the group leaders $z^1,...,z^T$ are assumed to be latent (as are $S^1,...,S^T$) since in many applications only the groups themselves are observable.

\begin{center}
\begin{figure}[!htb]
\begin{tikzpicture}

\tikzstyle{main}=[circle, minimum size = 5mm, thick, draw =black!80, node distance = 10mm]
\tikzstyle{connect}=[-latex, thick]
\tikzstyle{box}=[rectangle, draw=black!100]
  \node[box,draw=white!100] (Latent) {\textbf{     Latent}};
  \node[main] (L1) [right=of Latent] {$S^1$};
  \node[main] (L2) [right=of L1] {$S^2$};
  \node[main] (L3) [right=of L2] {$S^3$};
	\node[box,draw=white!100] (dots1) [right=of L3] {\textbf{$\boldsymbol{\cdots}$}};
  \node[main] (Lt) [right=of dots1] {$S^T$};
  \node[main,fill=black!10] (O1) [below=of L1] {$G^1$};
  \node[main,fill=black!10] (O2) [right=of O1,below=of L2] {$G^2$};
  \node[main,fill=black!10] (O3) [right=of O2,below=of L3] {$G^3$};
		\node[box,draw=white!100] (dots2) [right=of O3] {\textbf{$\boldsymbol{\cdots}$}};
  \node[main,fill=black!10] (Ot) [right=of dots2 ] {$G^T$};
	  \node[box,draw=white!100] (Observed) [left=of O1] {\textbf{     Observed}};

  \path (L1) edge [connect] (L2)
        (L2) edge [connect] (L3)
        (L3) edge [connect] (dots1)
        (dots1) edge [connect] (Lt);
  \path (O1) edge [connect] (O2)
        (O2) edge [connect] (O3)
				(O3) edge [connect] (dots2)
				(dots2) edge [connect] (Ot);
  \path (L1) edge [connect] (O1);
  \path (L2) edge [connect] (O2);
  \path (L3) edge [connect] (O3);
  \path (Lt) edge [connect] (Ot);

\end{tikzpicture}
\caption{A Bayesian network representing the temporal-dependent hub model. Nodes with dark colors indicate the observed variables.}
	\label{F:Bayesian}

\end{figure}
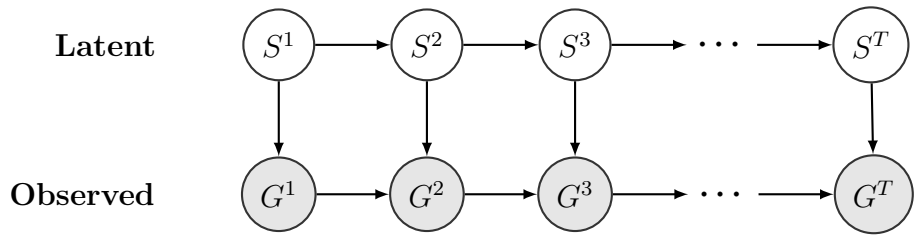
\end{center}

\section{Model fitting}\label{sec:alg}

In this section, we propose an algorithm to find the maximum likelihood estimators (MLEs) for $\alpha,\beta,\gamma,u$ and $\theta$. With $S$ being the latent variables, an expectation-maximization (EM) algorithm will be used for this problem. The EM algorithm maximizes the marginal likelihood of the observed data, which in our case is $G$, by iteratively applying an E-step and an M-step. 

Let $\alpha^{\textnormal{old}},\beta^{\textnormal{old}},\gamma^{\textnormal{old}},u^{\textnormal{old}}$ and $\theta^{\textnormal{old}}$ be the estimates in the current iteration. In the E-step, we calculate the conditional expectation of the complete log-likelihood given $G$ under the current estimate. That is, 
\begin{align*}
Q \overset{\Delta}{=} Q(\alpha,\beta,\gamma,u,\theta| \alpha^{\textnormal{old}},\beta^{\textnormal{old}},\gamma^{\textnormal{old}},u^{\textnormal{old}},\theta^{\textnormal{old}})= \mathbb{E}_{\alpha^{\textnormal{old}},\beta^{\textnormal{old}},\gamma^{\textnormal{old}},u^{\textnormal{old}},\theta^{\textnormal{old}}} \left [ \log \mathbb{P} (S,G) | G \right ].
\end{align*}
In the M-step, we maximize this conditional expectation with respect to the unknown parameters. That is,
\begin{align*}
(\alpha^{\textnormal{new}},\beta^{\textnormal{new}},\gamma^{\textnormal{new}},u^{\textnormal{new}},\theta^{\textnormal{new}} ) = \argmax_{\alpha,\beta,\gamma,u,\theta} \,\, Q(\alpha,\beta,\gamma,u,\theta| \alpha^{\textnormal{old}},\beta^{\textnormal{old}},\gamma^{\textnormal{old}},u^{\textnormal{old}},\theta^{\textnormal{old}}).
\end{align*}

It has been proved by \citet{wu1983convergence} that the EM algorithm converges to a local maximizer of the marginal likelihood. (Refer to \citet{McLachlan08} for a comprehensive introduction to this algorithm). We now give details of the two steps in our context. 
\subsection{E-step}\label{sec:E-step}
Since the complete log-likelihood $\log \mathbb{P} (S,G)$ is a linear function of $S_i^t \,\, (t=1,...,T;i=1,...,n)$ and $S_i^{t}S_j^{t-1} \,\,(t=2,...,T;i=1,...,n;j=1,...,n) $, the computation of its conditional expectation is equivalent to calculating $\mathbb{P}(S_i^t=1|G)$ and $\mathbb{P}(S_i^t=1,S_j^{t-1}=1|G)$. From now on, all conditional probabilities are defined under the current estimates.

A brute-force calculation of these probabilities, such as  
\begin{align*}
\mathbb{P}(S_i^t=1|G) = \mathbb{P} (z^t=i|G)=\frac{\sum_{z^1}\cdots \sum_{z^{t-1}}\sum_{z^{t+1}}\cdots\sum_{z^t} \mathbb{P}(z^1,...,z^{t-1},z^t=i,z^{t+1},...,z^T,G)}{\mathbb{P}(G)},
\end{align*}
is infeasible since the numerator involves a sum of $n^{T-1}$ terms. This is because $G^1,...,G^T$ are not independent according to our model. An efficient algorithm is needed for all practical purposes. 

The temporal-dependent hub model is similar to the hidden Markov model (HMM) (Figure \ref{F:Bayesian}). A polynomial-time algorithm for this model, called the forward-backward algorithm, was developed for computing the conditional probabilities. See \citet{smyth1997probabilistic,ghahramani2001introduction} for tutorials on HMMs and this algorithm. 

In the HMM, the observed variable at time $t$ only depends on the corresponding hidden state. But in our model, $G^t$ depends on both the current leader $z^t$ and the previous group $G^{t-1}$. We develop a new forward-backward algorithm for our model, which has more steps than the original algorithm but is also polynomial-time. We describe the algorithm here (see the Appendix for detailed derivation and justification). 

Define $a=[a_i^t],b=[b_i^t]$ and $c=[c_i^t]$ as $T\times n$ matrices. These matrices are computed by the following recursive procedures. 
\begin{align*}
a_i^1 & = \mathbb{P}(z^1=i,G^1) \quad (i=1,...,n). \\
a_i^t & =\sum_{k=1}^n a^{t-1}_k \Phi_{ik} \mathbb{P}(G^t|z^t=i,G^{t-1}) \quad (t=2,...,T;i=1,...,n). \\
b_i^T & = 1 \quad (i=1,...,n). \\
b_i^t & =\sum_{k=1}^n b^{t+1}_k \Phi_{ki} \mathbb{P}(G^{t+1}|z^{t+1}=k, G^t) \quad (t=T-1,...,1;i=1,...,n). \\
c_i^T & = \mathbb{P}(G^T|z^{T}=i,G^{T-1}) \quad (i=1,...,n). \\
c_i^t & = \sum_{k=1}^n c^{t+1}_k \Phi_{ki} \mathbb{P}(G^t|z^t=i,G^{t-1})  \quad (t=T-1,...,2;i=1,...,n). 
\end{align*}
 The matrices $a$, $b$ and $c$ should not be confused with the matrices $A$, $B$ and $C$ introduced in Section \ref{sec:Model}. The symbols are case-sensitive throughout the paper.

With these quantities,
\begin{align*}
\mathbb{P}(S_i^t=1|G) & = \frac{a_i^t b_i^t}{\sum_k a_k^t b_k^t} \quad (t=2,...,T;i=1,...,n). \\
\mathbb{P}(S_i^t=1,S_j^{t-1}=1|G) & = \frac{a_j^{t-1} \Phi_{ij} c_i^t}{\sum_{kl} a_l^{t-1} \Phi_{kl} c_k^t} \quad (t=2,...,T;i=1,...,n;j=1,...,n).
\end{align*} 
The complexity of this algorithm is $O(Tn^2)$. 
 
Note that the first row of $c$ is undefined but also unused. Also note that the elements of $a,b$ and $c$ will quickly vanish as the recursions progress. Therefore, we renormalize each row to sum to one at each step. It can easily be verified that this normalization does not affect the conditional probabilities. Finally, we emphasize that this algorithm gives the exact values of the conditional probabilities in a fixed number of steps -- i.e., it is not an approximate or iterative method.

\subsection{M-step}\label{sec:coord}
The M-step is somewhat routine compared to the E-step. First, it is clear that $\{ \alpha,u \}$ and $\{\beta,\gamma,\theta \}$ can be handled separately. 

We apply the coordinate ascent method (see \citet{Boyd04} for a comprehensive introduction) to iteratively update $\alpha$ and $u$, as well as $\beta,\gamma$ and $\theta$. Since the complete log-likelihood is concave and so is $Q$, coordinate ascent can guarantee a global maximizer. 

At each step, we optimize the log-likelihood over parameter one by one with the other parameters being fixed. The procedure is repeated until convergence. At each step, we use the standard Newton-Raphson method to solve each individual optimization problem. Specifically, for a parameter $\phi$ (here $\phi$ can represent $\alpha$, $\beta$, $\gamma$, $u_i$ or $\theta_{ij} \,\, (i < j)$ ), the estimate at $(m+1)$-th iteration is updated by the following formula given its estimate at $m$-th iteration: 
\begin{align*}
\hat{\phi}_{m+1}=\hat{\phi}_m-\left ( \frac{\partial^2 Q}{\partial \phi^2 }  \bigg|_{\phi=\hat{\phi}_m} \right )^{-1} \left ( \frac{\partial Q}{\partial \phi }  \bigg|_{\phi=\hat{\phi}_m}\right ). 
\end{align*}
The calculation of these derivatives is straightforward but tedious, so we provide the details in the Appendix. 

As shown in Section \ref{generating}, the model is not identifiable with respect to $u$. A standard solution to this problem is to set some $u_i \equiv 0$. But it does not work for our case. This is because for small data sets, some $\hat{\rho}_i$ estimated by the EM algorithm may be zero, implying that $v_i$ never became the leader. Furthermore, these zero $\rho_i$ cannot be predetermined since the leaders are unobserved. We observe that without constraint on $u_i$, the algorithm converges to different $\hat{u}$ with different initial values, but the corresponding $\hat{\rho}$ will be the same. Therefore, identifiability is not an issue for model fitting.


\subsection{Initial value}
As with many optimization algorithms, the EM algorithm is not guaranteed to find the global maximizer. Ideally, one should use multiple random initial values and find the best solution by comparing the marginal likelihoods $\mathbb{P}(G)$ under the corresponding estimates. 

In principle, $\mathbb{P}(G)$ can be computed by $\sum_k a_k^t b_k^t$, as shown in Section \ref{sec:E-step}. But the marginal likelihood vanishes quickly, even with a moderate $T$. Note that we cannot renormalize $a$ and $b$ for the purpose of computing $\mathbb{P}(G)$. 

Therefore, we use the half weight index \citep{Dice45,Cairns87} as the initial value of $A$, which is defined by
\begin{equation*}
H_{ij} = \frac{2\sum_t G_i^{t} G_j^{t}}{\sum_t G_i^{t}+\sum_t G_j^{t}}.
\end{equation*}
This measure estimates the conditional probability that two nodes co-occur given that one of them is observed, which is a reasonable initial guess of the strength of links. Furthermore, we use zero for the initial values of $\alpha, \beta$ and $\gamma$, and $\sum_{t} G_i^t/T $ for the initial value of $\rho_i$.

\section{Simulation studies}\label{sec:simu}
In all simulation studies, we fix the size of the network to be $n=50$ and set $\beta=3$ and $\gamma=-1$. We generate $u_i$ as independently and identically distributed variables with $N(0,2)$ and $\rho_i= u_i/\sum_k u_k $. The parameters $\theta_{ij}\,\, (i<j)$ are generated independently with $N(-2,1)$. We generate $\theta_{ij}$ in this way to control the average link density of the network ($\approx$ 0.12), which is more realistic than a symmetric setting, i.e., $\theta_{ij} \sim N(0,1)$. For clarification, we will not use the prior information on $u$ and $\theta$ in our estimating procedure. That is, we still treat $u$ and $\theta$ as unknown fixed parameters in the algorithm. We generate them as random variables for the whole purpose of adding more variations to the parameter setup in our study. 

We consider three levels of $\alpha=\log((n-1)/2),\log(n-1),\log(2(n-1))$, which correspond to a leader from the previous group remaining unchanged in the current group with probabilities $1/3,1/2,2/3$ on average. For each $\alpha$ we try five different sample sizes, $T=1000,1500, 2000, 2500$ and 3000. 

\begin{figure}[!ht]
	\begin{center}

	\includegraphics[width=6 in]{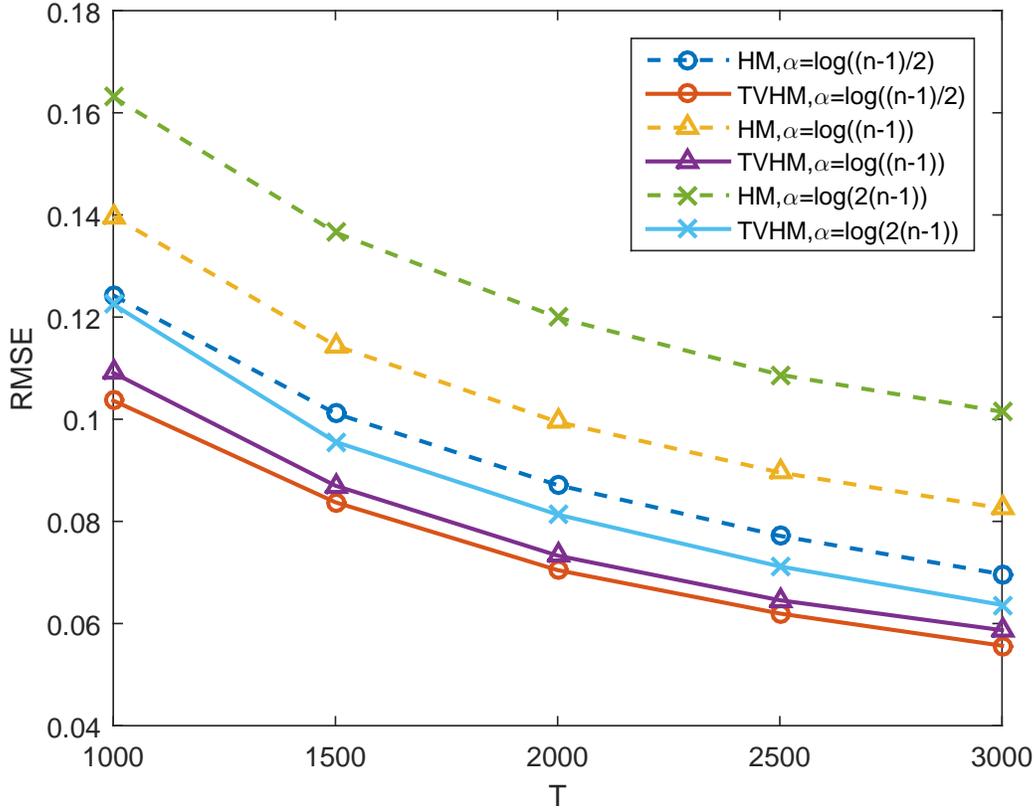}

	\caption{Average RMSEs (across 100 replicates) for the estimated $A$ from the classical hub model (HM) and the temporal-dependent hub model (TDHM). }
	\label{F:sim}
	\end{center}
\end{figure}

\begin{table}[!ht]
	\caption{Average RMSEs for the estimated $A$ from the classical hub model (HM) and the temporal-dependent hub model (TDHM). Standard deviations $\times 10^3$ in parentheses.}
	\label{Table:simulation2}
	\begin{center}
	\begin{tabular}{|c | cc | c c | cc |}
		\hline
		& \multicolumn{2}{c|}{$\alpha=\log((n-1)/2) $} & \multicolumn{2}{c|}{$\alpha=\log(n-1) $} & \multicolumn{2}{c|}{$\alpha=\log(2(n-1))$} \\ 
		\hline
		$T$ & HM & TDHM & HM & TDHM & HM & TDHM  \\
		\hline
1000 & 0.124	(7)	&	0.104	(6)	&	0.140	(8)	&	0.109	(7)	&	0.163	(12)	&	0.122	(10)	\\
1500 & 0.101	(7)	&	0.084	(6)	&	0.114	(8)	&	0.087	(6)	&	0.137	(9)	&	0.096	(9)	\\
2000 & 0.087	(5)	&	0.070	(5)	&	0.099	(6)	&	0.073	(5)	&	0.120	(8)	&	0.081	(8)	\\
2500 & 0.077	(5)	&	0.062	(4)	&	0.090	(5)	&	0.065	(5)	&	0.109	(7)	&	0.071	(7)	\\
3000 & 0.070	(4)	&	0.056	(4)	&	0.083	(4)	&	0.059	(4)	&	0.101	(6)	&	0.064	(6)	\\

		\hline 

	\end{tabular}
	\end{center}
\end{table}
Figure \ref{F:sim} shows the average root of mean squared errors (RMSEs) for the estimated $A$ over 100 replicates. For each simulation, we compare two methods, the classical hub model (HM) and the temporal-dependent hub model (TDHM). We assume the leaders are unknown under both models. Table \ref{Table:simulation2} provides the same information (with standard deviations) in numerical form. 


From Figure \ref{F:sim} and Table \ref{Table:simulation2}, our first observation is simply that the RMSEs decrease as sample sizes increase, which is consistent with common sense in statistics. 

Second, the RMSEs for all the parameters increase as $\alpha$ increases. This phenomenon can be interpreted as follows: with a larger value of $\alpha$, the correlation between adjacent groups becomes stronger and hence the effective sample size becomes smaller. The ratio of the sample size to the number of parameters decreases with $\alpha$, which makes inferences more difficult.

Third, the temporal-dependent hub model always outperforms the hub model. Moreover, the discrepancy between the temporal-dependent hub model estimates and the corresponding hub model estimates becomes larger as $\alpha$ increase. This is because the behavior of the temporal-dependent model deviates more from the classical hub model as $\alpha$ increases. 

The standard deviations and means show a similar trend. That is, the standard deviations decrease as $n$ increases and $\alpha$ decreases. The standard deviations for the temporal-dependent hub model estimates are comparable to or slightly smaller than those of the hub model estimates.


\section{A data example of group dynamics in chimpanzees}\label{sec:data}

Behavioral ecologists become increasingly interested in using social network analysis to understand social organization and animal behavior \citep{Bejder98,Whitehead2008,croft2011hypothesis,farine2012}. The social relationships are usually inferred by using certain association metrics (e.g., the half weight index) on grouped data. As indicated in the Introduction however, it is unclear how the inferred network relates to the observed groups without specifying a model. 

In this section, we study a data set of groups formed by chimpanzees by the temporal-dependent hub model. This data set is compiled from the results of the Kibale Chimpanzee Project, which is a long-term field study of the behavior, ecology and physiology of wild chimpanzees in the Kanyawara region of Kibale National Park, southwestern Uganda (https://kibalechimpanzees.wordpress.com/). 

Our analysis focuses on grouping behavior. We analyze the grouped data collected from January 1, 2009 to June 30, 2009 \citep{15390_2011}. The group identification was taken at 1 p.m. daily during this time period. If there is no group observed at 1 p.m. for a given day, it is not included in the data. Only one group is observed at 1 p.m. in 75.29\% of the remaining days over this period of six months. In the other days, multiple groups (usually two) are observed at 1 p.m. For these cases, we keep the group that has the most overlap with the previous group in our analysis. We use the Jaccard index to measure the overlap between two groups $G^{t-1}$ and $G^{t}$,
\begin{align*}
J(G^{t-1},G^{t})=\frac{\sum_{j=1}^n G^{t-1}_j G^{t}_j }{\sum_{j=1}^n [ n-(1-G^{t-1}_j)(1-G^{t}_j) ]},
\end{align*}
where the numerator is the size of the intersection of two groups and the denominator is the size of their union. One may refer to \citet{liben2007link} for an introduction to this measure.

Moreover, five chimpanzees never appear in any group and thus are removed. After the preprocessing, the data set contains 170 groups with 40 chimpanzees.

\begin{figure}[!ht]
  \hspace{1.5 in}
	\includegraphics[width=3.5 in]{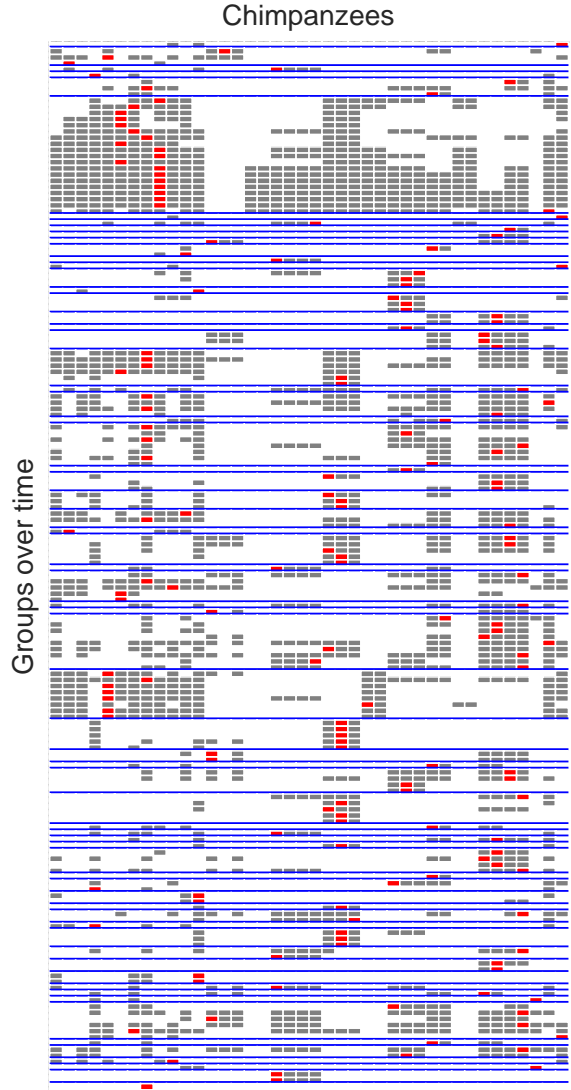}
	\vspace{-0.5 in}
	\caption{Visualization of the chimpanzee data. The 170 rows represent the groups over time and the 40 columns represent chimpanzees. Gray indicates the presence of the group membership. Red indicates the inferred group leaders. The blue lines separate the inferred segments.}
	\label{F:monkey_data}
\end{figure}

Figure \ref{F:monkey_data} illustrates the data set in grayscale with rows representing the groups over time and columns representing the chimpanzees. Black indicates $G^{t}_i=1$ at location $(t,i)$ while white indicates $G^{t}_i=0$. The pattern in Figure \ref{F:monkey_data} clearly demonstrates the existence of dependency between groups. 

Figure \ref{F:monkey_data} also shows the inferred grouped leaders indicated in red with the inferred segments separated by blue lines. By the inferred grouped leader for $G^t$, we mean that the chimpanzee with the highest posterior probability will be the leader given $G^t$. As shown in Figure \ref{F:monkey_data}, the leaders retain a certain level of stability, which is consistent with the estimates of $\alpha$ ($=1.7291$). Also, recall that by our definition, a new segment starts if the current leader is not within the previous group. From Figure \ref{F:monkey_data}, the inferred segments are coincident with the visualization of the data set. 
 
The estimated values of the adjustment factors, $\hat{\beta}=2.5703$ and $\hat{\gamma}=-0.1922$. The magnitude of $\hat{\beta}$ is larger than that of $\hat{\gamma}$, which suggests individuals have a stronger tendency to join a group than leave a group. In other words, the groups may start with small size and grow larger over time. This phenomenon is shown (Figure \ref{F:monkey_data}).

\begin{figure}[!ht]
\begin{center}
  \centerline{\hfill\makebox[2.8 in]{(a) HM}\hfill\makebox[2.8 in]{(b) TDHM}\hfill}
  \centerline{\hfill
    \includegraphics[width=2.8 in, height=2.8 in]{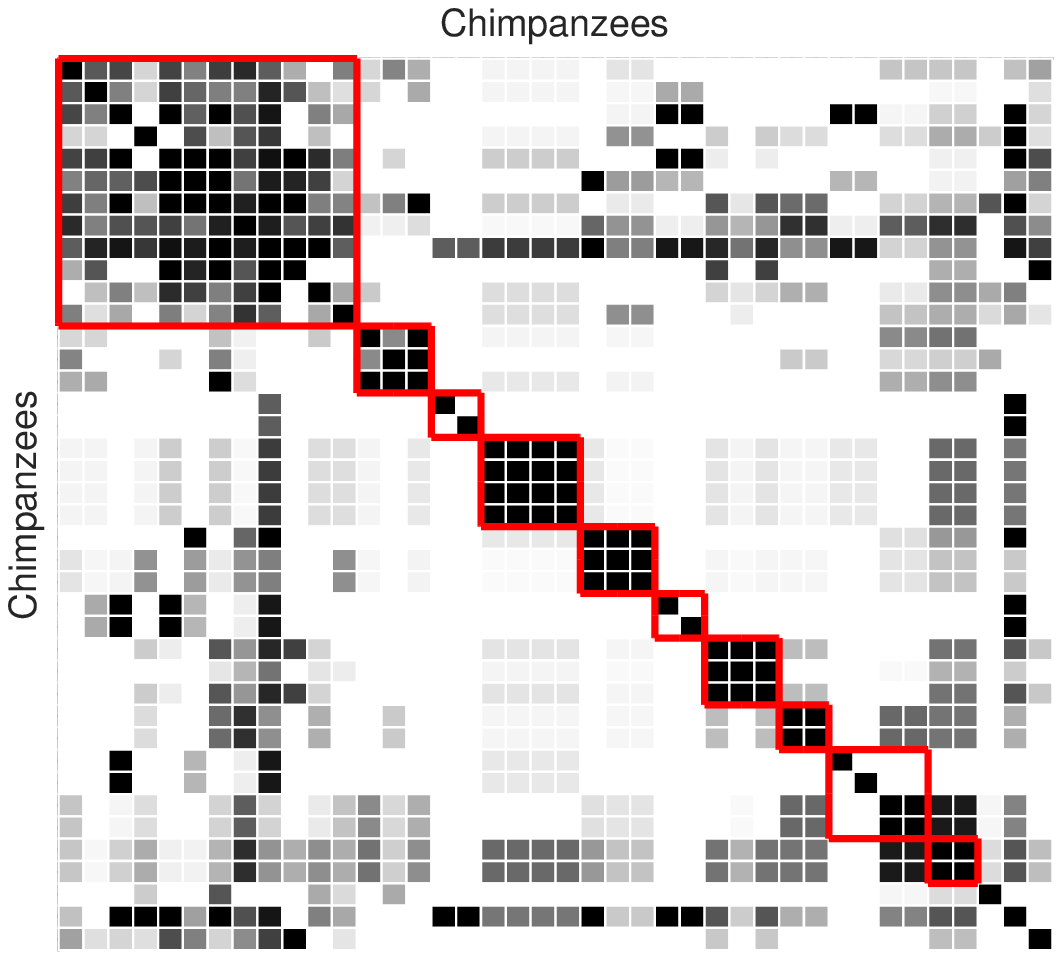}
    \hfill
    \includegraphics[width=3 in, height=2.8 in]{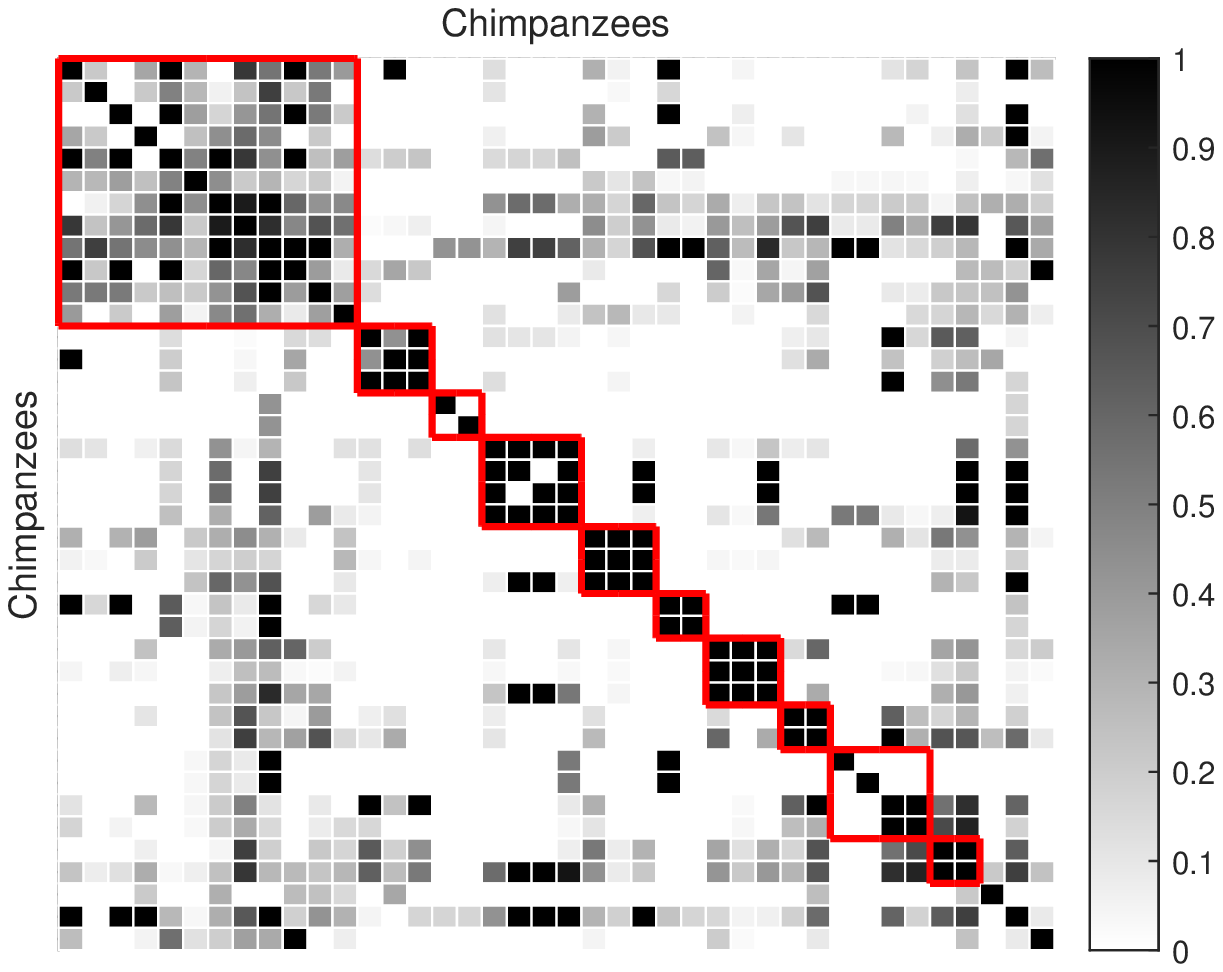}
    \hfill}
			\vspace{-0.2 in}

			\caption{Grayscale plot for the estimated adjacency matrices from the chimpanzee data by the classical hub model and the temporal-dependent hub model. The rows and the columns represent the 40 chimpanzees.  Darker colors indicate stronger relationships. The red blocks indicate the biological clusters of chimpanzees.}
\label{F:monkey_result}
\end{center}
\end{figure}

Figure \ref{F:monkey_result} shows the result of estimated adjacency matrices by the classical hub model and the temporal-dependent hub model. As in the previous figure, the darker color indicates a higher value of $\hat{A}_{ij}$. The red blocks indicate clusters of chimpanzees in a biological sense. The first cluster consists of 12 adult males and each of the other nine clusters consists of an adult female and its children. From the estimates by both the classical hub model and the temporal-dependent hub model, there are strong connections within these biological clusters. Both estimates suggest that in this data set of chimpanzees, adult males usually do activities together but females usually stay with their children.

 The two estimated adjacency matrices are different, however. Generally speaking, without properly considering the temporal-dependence between groups, the estimates of the relationships between individuals by the classical hub model can be biased. That is, an individual may choose to stay within or out of a group not solely based on its relationship with the group leader but also because of the inertia. The overall graph density $(= 0.2286)$ of the estimated network by the classical hub model is larger than the corresponding value $(= 0.1973)$ of the temporal-dependent hub model. This is consistent with the fact that the magnitude of $\hat{\beta}$ is larger than the magnitude of $\hat{\gamma}$. Since the classical hub model does not incorporate the adjustment factors, bias is introduced to certain $\hat{A}_{ij}$ so the model can match the overall frequency of occurrences for the individuals.

\begin{figure}[!ht]
\begin{center}
\twoImages{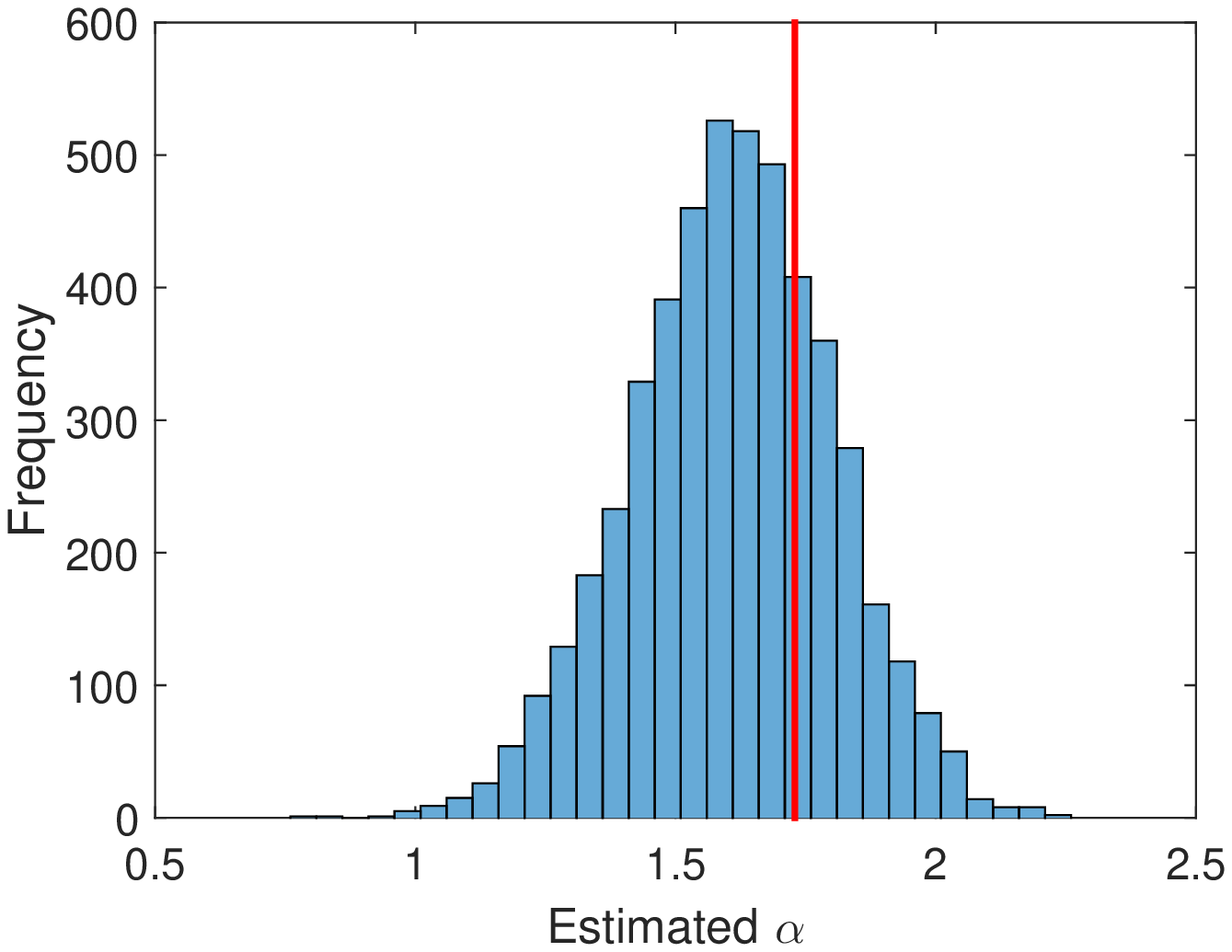}{2.8 in}{}
{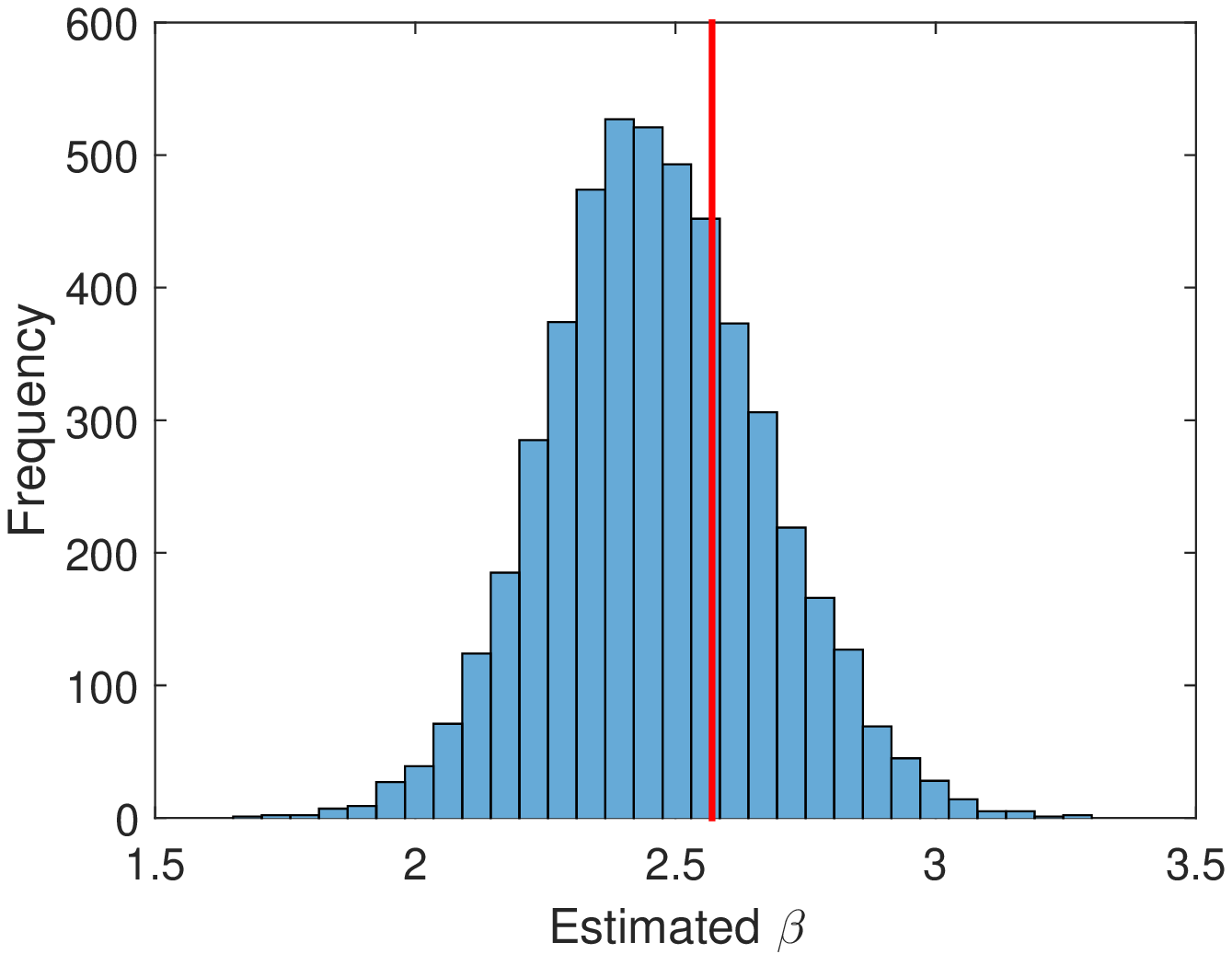}{2.8 in}{}
\makebox[2.8 in]{} \\
\includegraphics[width=2.8 in]{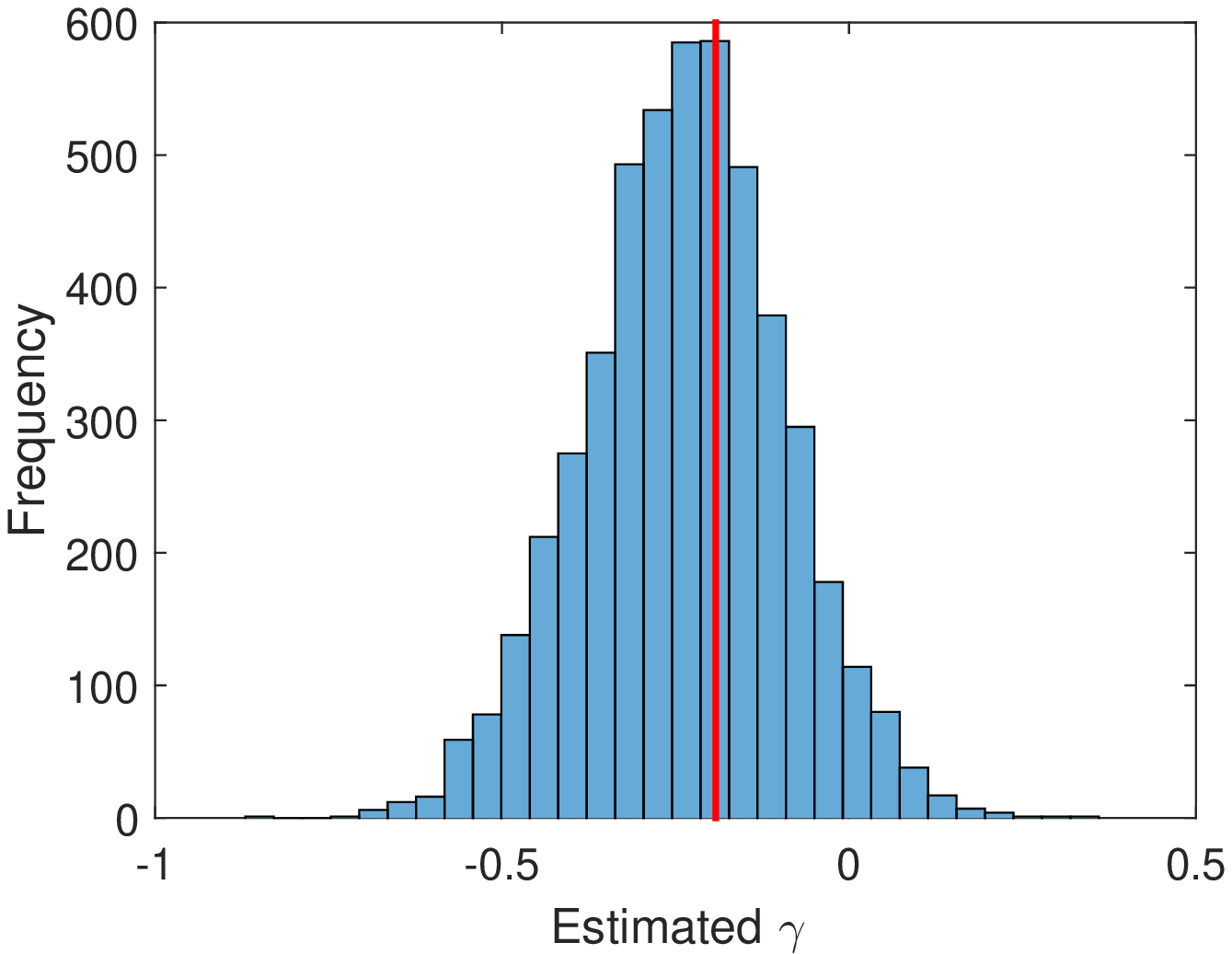}
\caption{Histograms of estimates from parametric bootstrap samples. Red lines indicate the estimated values from the original data set.}
\label{F:hist}
\end{center}
\end{figure}

The significance of $\alpha$, $\beta$ and $\gamma$ is tested by the parametric bootstrap method \citep{efron1994introduction}. Specifically, we generate 5000 independent data sets from the fitted temporal-dependent hub model to the original data and compute the MLEs for each simulated data set. The parametric bootstrap was applied to HMMs and showed a good performance \citep{visser2000confidence}.  Figure \ref{F:hist} shows the histograms of the MLEs for $\alpha$, $\beta$ and $\gamma$. The 95\% bootstrap confidence intervals for $\alpha$, $\beta$ and $\gamma$ are (1.2177, 1.9774), (2.0710, 2.8944) and (-0.5208, 0.0410), which shows that the effects of $\alpha$ and $\beta$ are significant while $\gamma$ is not at the 0.05 significance level. This further supports the observation in the previous paragraph -- chimpanzees have a stronger tendency to join a group than to leave a group in this data set. 

\section{Summary and discussion} \label{sec:con}
In this article, we generalize the idea of the hub model and propose a novel model for temporal-dependent grouped data. This new model allows for dependency between groups. Specifically, the group leaders follow a Markov chain and a group is either a transformation of the previous group or a new start, depending on whether the current leader is within the previous group. An EM algorithm is applied to this model with a polynomial-time algorithm being developed for the E-step. 

The setup of our model is different from some work on estimating time-varying networks by graphical models, e.g., \citet{kolar2010estimating} for discrete data and \citet{Zhou2010} for continuous data. These papers assume that the observations are independent and the latent network changes smoothly or is piecewise constant. In this paper, we instead focus on the dependence between groups. Ideally, both aspects -- dependence between groups and changes in networks -- need to be considered in modeling. This should be plausible when the sample size, i.e., the number of observed groups, is large. When the sample size is moderate (as in the chimpanzee data set), the length of the time interval between observation plays a key role in determining which aspect is more important. If the time interval is short, then dependence between groups is significant but the changes in the latent network are likely to be minor since the overall time window is not long. On the contrary, the dependence between groups becomes weak when the time interval is long. 

For future work, we plan to study the time-varying effect on the latent network for the grouped data. When changepoints exist, a single network cannot accurately represent the link strengths in different time windows. A change-point analysis for temporal-dependent grouped data is an intriguing research topic. Alternative approaches may be based on penalizing the difference between networks at adjacent time points, although careful investigation is required to determine how tractable these methods are for temporal-dependent groups. 

In addition to time-varying networks, the temporal-dependent hub model can also be extended in the following directions: first, a group may contain zero or multiple hubs. Second, multiple groups may exist at the same time (with some of these groups being unobserved). These generalizations however will significantly increase model complexity. Therefore, the total number of possible leaders needs to be limited. A method by the author and a collaborator \citep{weko2017penalized} was proposed to reduce this upper bound. More test-based and penalization methods are under development. 

Furthermore, we also plan to investigate the theoretical properties of the proposed model. When the size of the network is fixed and the number of observed groups goes to infinity, the theoretical properties of the MLE may be studied via a standard theory of the Markov chain. The case that the size of the network also goes to infinity is more intriguing but more complicated since the number of parameters diverges.

\section*{Acknowledgments}
This research was supported by NSF Grant DMS 1513004. We thank Dr. Richard Wrangham for sharing the research results of the Kibale Chimpanzee Project. We thank Dr. Charles Weko for compiling the results from the chimpanzee project and preparing the data set.

\appendix

\section{Forward-backward algorithm in the E-steps}\label{app1}
We derive the forward-backward algorithm for the temporal-dependent hub model introduced in Section \ref{sec:E-step}. Before proceeding, we state two propositions of Bayesian networks. These results (or the equivalent forms) can be found in a standard textbook or tutorial on Bayesian networks, for example, \citet{jordan1999introduction}. Here we follow \cite{ghahramani2001introduction}.
\begin{prop}\label{thm1}
Each node is conditionally independent from its non-descendents given its parents. Here node $X$ is a parent of another node $Y$ if there is a directed arc from $X$ to $Y$ and if so, $Y$ is a child of $X$. The descendents of a node are its children, children's children, etc. 
\end{prop}
\begin{prop}\label{thm2}
Two disjoint sets of nodes $\mathcal{A}$ and $\mathcal{B}$ are conditionally independent given another set $\mathcal{C}$, if on every \textit{undirected} path between a node in $\mathcal{A}$ and a node in $\mathcal{B}$, there is a node $X$ in $\mathcal{C}$ that is not a child of both the previous and following nodes in the path.   
\end{prop}

Define $G^{s:t}$ as a collection of groups from time $s$ to time $t$. 

Let $a_i^t=\mathbb{P}(z^t=i,G^{1:t})$. Then,
\begin{align*}
a_i^t & =  \sum_{k=1}^n\mathbb{P}(z^t=i,z^{t-1}=k,G^{1:t-1},G^t) \\
      & =  \sum_{k=1}^n\mathbb{P}(z^{t-1}=k,G^{1:t-1})\mathbb{P}(z^t=i|z^{t-1}=k,G^{1:t-1})\mathbb{P}(G^t|z^t=i,z^{t-1}=k,G^{1:t-1}) \\
			& = \sum_{k=1}^n a^{t-1}_k \Phi_{ik} \mathbb{P}(G^t|z^t=i,G^{t-1}).
\end{align*}
The last equation holds by Proposition \ref{thm1}. 

Similarly, let $b_i^t=\mathbb{P}(G^{t+1:T}|z^t=i, G^t)$. Then,
\begin{align*}
b_i^t & = \sum_{k=1}^n \mathbb{P}(z^{t+1}=k, G^{t+1},G^{t+2:T}|z^t=i, G^t) \\
      & = \sum_{k=1}^n \mathbb{P}(G^{t+2:T}|z^{t+1}=k,G^{t+1}, z^t=i, G^t)\mathbb{P}(z^{t+1}=k|z^t=i, G^t)\mathbb{P}(G^{t+1}|z^{t+1}=k,z^t=i, G^t) \\
			& = \sum_{k=1}^n b^{t+1}_k \Phi_{ki} \mathbb{P}(G^{t+1}|z^{t+1}=k, G^t).
\end{align*}
In the last equation, $\mathbb{P}(G^{t+2:T}|z^{t+1}=k,G^{t+1}, z^t=i, G^t)$ holds by Proposition \ref{thm2}. This is because a path from $\{G^{t+2:T}\}$ to $\{ z^t, G^t\}$ must pass $z^{t+1}$ or $G^{t+1}$. If it only passes one of these two variables, then we can take that variable as $X$ in Proposition \ref{thm2}. If it passes both, then take $z^{t+1}$ as $X$. The rest of the last equation holds by \ref{thm1}. 

The computation of $a$ and $b$ is essentially the same as in the classical forward-backward algorithm for the HMM with minor modifications. Unlike the HMM, the dependence between the current and the previous groups requires another quantity $c$. 
 
Let $c_i^t=\mathbb{P}(G^{t:T}|z^t=i, G^{t-1}).$
\begin{align*}
c_i^t & = \sum_{k=1}^n \mathbb{P}(z^{t+1}=k,G^{t+1:T},G^t|z^t=i,G^{t-1}) \\
      & = \sum_{k=1}^n \mathbb{P}(G^{t+1:T}|z^{t+1}=k,G^t,z^t=i,G^{t-1}) \mathbb{P}(z^{t+1}=k|z^t=i,G^{t-1}) \mathbb{P}(G^t|z^{t+1}=k,z^t=i,G^{t-1}) \\
			& = \sum_{k=1}^n c^{t+1}_k\Phi_{ki} \mathbb{P}(G^t|z^t=i,G^{t-1}).
\end{align*}
The last equation can be justified by a similar argument as before.
 
Since
\begin{align*}
 & \mathbb{P}(z^t=i,G^{1:T}) \\
= & \mathbb{P}(z^t=i,G^{1:t})\mathbb{P}(G^{t+1:T}| z^t=i, G^{1:t}) \\
= & \mathbb{P}(z^t=i,G^{1:t})\mathbb{P}(G^{t+1:T}|z^t=i, G^t),
\end{align*}
\begin{align*}
\mathbb{P}(S_i^t=1|G) & = \frac{a_i^t b_i^t}{\sum_k a_k^t b_k^t}.
\end{align*}
Similarly,
\begin{align*}
 & \mathbb{P}(z^t=i,z^{t-1}=j,G^{1:T}) \\
= & \mathbb{P}(z^{t-1}=j,G^{1:t-1})\mathbb{P}(z^t=i|z^{t-1}=j,G^{1:t-1})\mathbb{P}(G^{t:T}|z^t=i,z^{t-1}=j,G^{1:t-1})  \\
= & a_j^{t-1} \Phi_{ij} \mathbb{P}(G^{t:T}|z^t=i,G^{t-1}) ,
\end{align*}
which implies
\begin{align*}
\mathbb{P}(S_i^t=1,S_j^{t-1}=1|G) & = \frac{a_j^{t-1} \Phi_{ij} c_i^t}{\sum_{kl} a_l^{t-1} \Phi_{kl} c_k^t}.
\end{align*}

\section{Derivatives of $Q$}\label{app2}
We give the first and second derivatives of $Q$ with respect to $\alpha, \beta, \gamma, u_i$ and $\theta_{ij}$, which are used in the coordinate ascent method introduced in Section \ref{sec:coord}. 

Define,
\begin{align*}
R_{i}^t &=\mathbb{P}(S_i^t=1|G), \\
V_{ij}  &=\sum_{t=2}^T \mathbb{P}(S_i^t=1,S_j^{t-1}=1|G), \\
D^1_{ij} &= R_i^{1} G_j^{1} +\sum_{t=2}^T R_i^{t} (1-G_i^{t-1}) G_j^{t}, \\
D^2_{ij} & =R_i^{1} (1-G_j^{1})+\sum_{t=2}^T R_i^{t}(1-G_i^{t-1}) (1-G_j^{t}), \\
D^3_{ij} & = \sum_{t=2}^T R_i^{t}G_i^{t-1} G_j^{t-1} G_j^{t}, \\
D^4_{ij} & = \sum_{t=2}^T R_i^{t}G_i^{t-1} G_j^{t-1} (1-G_j^{t}), \\
D^5_{ij} & = \sum_{t=2}^T R_i^{t}G_i^{t-1} (1-G_j^{t-1}) G_j^{t}, \\
D^6_{ij} & = \sum_{t=2}^T R_i^{t}G_i^{t-1} (1-G_j^{t-1}) (1-G_j^{t}).
\end{align*}
Therefore, 
\begin{align*}
Q  = & \sum_{i=1}^n R_i^{1}  \left [ u_i-\log \left \{ \sum_{k=1}^{n} \exp(u_k)   \right \} \right ] \\
 & + \sum_{i=1}^n V_{ii} \left [u_i+\alpha- \log \left \{ \sum_{k=1}^{n} \exp(u_k+\alpha I(k=i))   \right \} \right ] \\
 & +  \sum_{i=1}^n \sum_{j \neq i}V_{ij} \left [u_i- \log \left \{ \sum_{k=1}^{n} \exp(u_k+\alpha I(k=j))   \right \} \right ] \\
 & + \sum_{ij}     \left [  D^1_{ij} \log \frac{e^{\theta_{ij}}}{1+e^{\theta_{ij}}} +  D^2_{ij} \log \frac{1}{1+e^{\theta_{ij}}}  \right . \\
 & \quad \quad + D^3 _{ij} \log \frac{e^{\theta_{ij}+\beta}}{1+e^{\theta_{ij}+\beta}} +D^4 _{ij} \log \frac{1}{1+e^{\theta_{ij}+\beta}} \\
 & \quad \quad  + \left . D^5 _{ij} \log \frac{e^{\theta_{ij}+\gamma}}{1+e^{\theta_{ij}+\gamma}} +D^6 _{ij} \log \frac{1}{1+e^{\theta_{ij}+\gamma}} \right ] . 
\end{align*}
The first and second order derivatives are given as follows,
\begin{align*}
\frac{\partial Q }{\partial \alpha } = &  \sum_{i=1}^nV_{ii} + \sum_{j=1}^n \left [\sum_{i=1}^n  V_{ij} \right ]\left [- \frac{\exp(u_j+\alpha)}{\sum_{k=1}^{n} \exp(u_k+\alpha I(k=j)) } \right ], \\
\frac{\partial^2 Q }{\partial \alpha^2}= 
&  \sum_{j=1}^n \left [\sum_{i=1}^n V_{ij} \right ]\left [- \frac{\exp(u_j+\alpha)\sum_{k=1}^{n} \exp(u_k+\alpha I(k=j))-\exp(u_j+\alpha)\exp(u_j+\alpha)}{(\sum_{k=1}^{n} \exp(u_k+\alpha I(k=j)))^2 }\right ], \\
\frac{\partial Q }{\partial u_r } = & R_r^{1} + \left [ \sum_{i=1}^n R_i^{1} \right ] \left [-\frac{\exp(u_r)}{\sum_{k=1}^{n} \exp(u_k) } \right ], \\
\frac{\partial^2 Q }{\partial u_r^2 } = &  \left [\sum_{i=1}^n R_i^{1} \right ]\left [-\frac{\exp(u_r)\sum_{k=1}^{n} \exp(u_k)-\exp(u_r)\exp(u_r)}{(\sum_{k=1}^{n} \exp(u_k))^2} \right ] \\
 & +\sum_{j=1}^n \left[\sum_{i=1}^n V_{ij} \right ] \left [ -\frac{\exp(u_r+\alpha I(r=j))\sum_{k=1}^{n} \exp(u_k+\alpha I(k=j))-(\exp(u_r+\alpha I(r=j)))^2}{(\sum_{k=1}^{n} \exp(u_k+\alpha I(k=j)))^2 }   \right ] \\
& +\sum_{j=1}^n V_{rj} + \sum_{j=1}^n \left [\sum_{i=1}^n B_{ij} \right] \left [-\frac{\exp(u_r+\alpha I(r=j))}{\sum_{k=1}^{n} \exp(u_k+\alpha I(k=j)) }  \right ], \\
\frac{\partial Q}{\partial \beta}  = & \sum_{i \neq j}   D^3 _{ij}-( D^3 _{ij}+D^4_{ij}) \frac{e^{\theta_{ij}+\beta}}{1+e^{\theta_{ij}+\beta}}, \\
\frac{\partial^2 Q}{\partial \beta^2 }  = & -\sum_{i \neq j} ( D^3 _{ij}+D^4_{ij}) \frac{e^{\theta_{ij}+\beta}}{(1+e^{\theta_{ij}+\beta})^2}, \\
\frac{\partial Q}{\partial \gamma}  = & \sum_{i \neq j}   D^5 _{ij}-( D^5 _{ij}+D^6_{ij}) \frac{e^{\theta_{ij}+\gamma}}{1+e^{\theta_{ij}+\gamma}} , \\
\frac{\partial^2 Q}{\partial \gamma^2 }  = & -\sum_{i \neq j} ( D^5 _{ij}+D^6_{ij}) \frac{e^{\theta_{ij}+\gamma}}{(1+e^{\theta_{ij}+\gamma})^2}, \\
\frac{\partial Q}{\partial \theta_{ij}}  = &  (D^1_{ij}+D^1_{ji})-(D^1_{ij}+D^1_{ji}+D^2_{ij}+D^2_{ji}) \frac{e^{\theta_{ij}}}{1+e^{\theta_{ij}}} \\
 & + (D^3_{ij}+D^3_{ji})-(D^3_{ij}+D^3_{ji}+D^4_{ij}+D^4_{ji}) \frac{e^{\theta_{ij}+\beta}}{1+e^{\theta_{ij}+\beta}} \\
 & + (D^5_{ij}+D^5_{ji})-(D^5_{ij}+D^5_{ji}+D^6_{ij}+D^6_{ji}) \frac{e^{\theta_{ij}+\gamma}}{1+e^{\theta_{ij}+\gamma}}, \\
\frac{\partial^2 Q}{\partial \theta_{ij}^2} = & -(D^1_{ij}+D^1_{ji}+D^2_{ij}+D^2_{ji}) \frac{e^{\theta_{ij}}}{(1+e^{\theta_{ij}})^2} \\
 & -(D^3_{ij}+D^3_{ji}+D^4_{ij}+D^4_{ji}) \frac{e^{\theta_{ij}+\beta}}{(1+e^{\theta_{ij}+\beta})^2} \\
 & -(D^5_{ij}+D^5_{ji}+D^6_{ij}+D^6_{ji}) \frac{e^{\theta_{ij}+\gamma}}{(1+e^{\theta_{ij}+\gamma})^2}.
\end{align*}


\end{document}